\documentclass[prd,showpacs,amsmath,showkeys,twocolumn,floatfix]{revtex4} 
\usepackage{epsfig,dcolumn}
\usepackage{graphicx}
\usepackage[usenames]{color}
\usepackage{graphicx}
\usepackage{bm}
\usepackage{subfigure}
\usepackage{array} 

\def\k{{\bf k}}

\def\p{{\bf p}}

\def\lsim{\mathrel{\rlap{\lower4pt\hbox{\hskip1pt$\sim$}}
    \raise1pt\hbox{$<$}}}
\def\gsim{\mathrel{\rlap{\lower4pt\hbox{\hskip1pt$\sim$}}
    \raise1pt\hbox{$>$}}}

\begin{document}
\title{A model for QCD ground state with magnetic disorder} 
\author {Adam~P. Szczepaniak$^1$ and Hrayr.~H Matevosyan$^{1,2}$}
\affiliation{$^1$Physics Department and Nuclear Theory Center, \\
Indiana University, Bloomington, IN 47403\\
$^2$  Special Research Centre for the Subatomic Structure of Matter, \\
Department of Physics, Adelaide University, Adelaide, 5005, Australia}
\date{\today}

\begin{abstract}
We explore an ansatz for the QCD vacuum in the Coulomb gauge that  describes gauge field fluctuations in presence of a weakly-interacting gas of abelian monopoles.  Such magnetic disorder leads to long-range correlations which are manifested through the area law for the Wilson loop. In particular we 
focus on  the role of the residual monopole-monopole interactions in providing the mechanism for suppression of the gluon propagator at low-momenta  
 which also leads to low momentum enhancement in the ghost propagator.  

\end{abstract}

\maketitle

\section{\label{intro}Introduction}

A condensate of magnetic monopoles screens chromoelectric field between fundamental color charges and leads to  formation of  flux tubes between static quarks. Consequently  condensation of magnetic  degrees of freedom in the QCD vacuum has been proposed as the  mechanism underlying  color confinement~\cite{m1,m2,m3,m4} and recently strong evidence for magnetic dominance has  emerged from lattice gauge simulations~\cite{l1,l2,l3,l4}.  In  QCD  monopoles emerge as solutions of classical equations of 
motion~\cite{Wu:1967vp}, albeit having infinite energy, 
  the underlying ultraviolet singularity is expected to be regularized by quantum fluctuations~\cite{Banks:1977cc,Diakonov:1999gg} and in any case should not play a role in the low energy domain.   Monopoles create vortices~\cite{l1,mvv1,mvv2} and all together lead to 
 percolation of monopole-antimonopole chains~\cite{cp1}. Even though there is ample evidence for presence of  magnetic domains it is   still an open issue which of the many possible monopole-vortex  geometries  dominates the QCD vacuum~\cite{l4}.  It  should be noted however, that vortex-like configurations of long monopole chains  are needed to achieve the confining scenario based between charges associated with the center of the gauge group as 
  expected for QCD~\cite{Ambjorn:1998qp}. 
 
Recently lattice simulations in both Landau~\cite{Cucchieri:2007md,Bogolubsky:2009dc, Bornyakov:2008yx,Cucchieri:2007rg,Cucchieri:2008fc,Sternbeck:2008na,Maas:2007uv,Dudal:2008sp,Boucaud:2009sd}  and Coulomb gauge~\cite{Langfeld:2004qs,Voigt:2007wd,Burgio:2008jr} have significantly advanced  our  knowledge on  the infrared (IR) properties of QCD Green's functions. And together with studies in the continuum~\cite{von Smekal:1997is,von Smekal:1997vx,Zwanziger:2001kw,Alkofer:2004it,Fischer:2008uz,Aguilar:2008xm,Szczepaniak:2001rg,Szczepaniak:2003ve,Feuchter:2004mk,Epple:2006hv,Epple:2007ut}  have reinvigorated discussion on  the role of  Green's functions in probing 
 the long-range properties of the QCD  vacuum~\cite{Papavassiliou:2009vk,Alkofer:2009vr}. 
 For example, in the Gribov-Zwanziger (GZ)   scenario~\cite{gz1,gz2,gz3,gz4}, confinement is related to presence of large field configurations near the boundary of the Gribov region, the 
  Gribov horizon. The  Gribov region is defined as the domain in the gauge field space that satisfies  a particular gauge condition. Within the Gribov region the  Faddeev-Popov (FP)  operator
   is positive and  it vanishes on the horizon.  Thus in the GZ scenario one expects that the   ({\it vev}) of the  inverse  of the FP  operator,  referred to as the ghost propagator, is IR enhanced.  
     On the other hand, the gluon propagator, which represents propagation of color charges is expected to be suppressed. The IR enhancement of the ghost propagator and suppression of the gluon propagator have  unambiguously  been seen in lattice simulations in both Landau and Coulomb gauges.  This is to be contrasted with some of the phenomenological  models 
 were it is  assumed that the   quark-antiquark potential originates from the nonperturbative gluon exchange. Thus a confining quark-antiquark potential 
   would necessitate an IR enhanced gluon propagator~\cite{Maris:1999nt, Maris:1997hd}. 
 
 In this paper  we explore the possibility that the low-momentum  suppression of the  gluon propagator  may not necessarily be related to confinement but to 
color  screening.  In the latter case  IR suppression of the gluon propagator implies that colored, physical gluons do not propagate and as follows from the GZ conjecture it would  
 not be the role of the gluon propagator but of  the IR enhanced ghost propagator to carry the distinct signatures  
  of  confinement.  It is a known phenomena that a dynamically generated gluon mass and IR suppression of the gluon propagator can emerge as a result of screening {\it e.g} induced by condensation  of magnetic domains, (vortices, monopoles) ~\cite{sc1,sc2}.
It can be well illustrated in models~\cite{l4}, {\it i.e} the Fradkin-Shenker model~\cite{Fradkin:1978dv} where the confined and Higgs phase are smoothly connected and there is also a smooth 
 transition in the gauge propagators~\cite{Chernodub:2004yv}.  Similarly  in the Landau gauge, solutions of QCD Dyson-Schwinger equations that display screening behavior have recently been studied in~\cite{Aguilar:2008xm}. 
 
 In the following section  we examine the effect of condensation of magnetic monopoles in the Coulomb gauge. In particular we study an ansatz for the vacuum wave functional that contains a weakly interacting gas of monopoles.   The recently studies of  various models for  the  vacuum wave 
 function~\cite{Szczepaniak:2001rg,Szczepaniak:2003ve,Feuchter:2004mk,Epple:2006hv,Epple:2007ut,Quandt:2010yq,Greensite:1979yn,Greensite:2009mi,Greensite:2010tm}  were shown to be quite successful phenomenologically.  In particular analytical calculations can  be done with an ansatz that describes the vacuum in terms of  gaussian fluctuations of the transverse,  vector  potential  around the zero-field configurations.  Since  such a simple vacuum does not explicitly contain   magnetic degrees of freedom it leads to a perimeter law falloff of  the {\it vev} of a Wilson loop.    Nevertheless  onset of confinement could be seen through the IR enhancement of the ghost  propagator   and the non-abelian Coulomb potential between static charges. This is because  gaussian ansatz can support  large fluctuations of the fields, which may be nearing   
 the boundary of the Gribov region. We begin with the construction  of our vacuum ansatz followed by a  discussion  of  the gluon and ghost propagators as well as the WIlson loop. A summary and outlook are  given in Section~\ref{s2}. 

       \section{  Model for the monopole dominated vacuum } 
       
  In the following we consider $SU(2)$  Coulomb gauge QCD,  in the Shr\"odinger picture represented  by the transverse field variables, $\nabla  \cdot A^a =0$, where $a=1\cdots N_C^2-1=3$ is the color index. The gaussian approximation to the vacuum  discussed earlier  is given by a  wave function  in the from 
         \begin{equation} 
         \Psi^w_T[A_T]  \propto  e^{-\frac{1}{2}\int dx dy A_T(x) \omega_T(x-y)  A_T(y) }.  \label{psit} 
            \end{equation} 
We will use the T(L) subscript to represent 
 components in the color algebra  transverse (longitudinal)  to a chosen direction, $w = w^a$, ($w^2=1$),  {\it i.e.} $A^{ia}_T \equiv A^{ia} - (w^b A^{ib}) w^a$, $A^{i}_L \equiv A^{ia} w^a$.  The  meaning of $w$ and the role of fields along $w$ will be discussed below.  The 2-point function (hereafter referred to as the  gluon propagator)  of color-transverse gluons is  then given by  (${\cal V}$ is the  three-dimensional  volume )
 \begin{equation} 
{\cal V}^{-1}  \langle A_T(k) A_T(-k) \rangle = \delta_T(k) \delta_T(w) D_T(k),  
 \end{equation} 
  where $\delta_T(n) = \delta_T^{ij}(n) = \delta_{ij} - n^i n^j/n^2$ is the transverse projector in three dimensions and  $D_T(k) = 1/(2\omega_T(k))$ where  $\omega_T(k)$ is the Fourier transform of 
    $\omega_T(x)$ that appears in Eq.~(\ref{psit}). Our ansatz wave function  absorbs  the  Coulomb gauge functional measure~\cite{Reinhardt:2004mm} {\it i.e.} 
   \begin{equation} 
 \int DA {\cal J}[A] \langle \Psi_{QCD}| A\rangle^2 \to \int DA  \langle \Psi_{Ansatz} |  A \rangle^2=1. 
 \end{equation} 
 With the gaussian wave function to (approximately) in order to  restrict field configurations to the inside of the Gribov volume where the   FP operator ${\cal J}$  is positive,  $\omega_T(k)$  should be 
  taken such that it diverges  in the IR limit  {\it i.e.} $\omega_T(k \to 0) \propto k^{1+\alpha_{G,T}}$  with $\alpha_{G,T} < -1$. For large momenta, the choice $\omega_T(k) \to k$ 
 makes  the vacuum wave function match that of the free theory.  
 
 The separation of transverse and longitudinal, (with respect to $w$) components 
   is motivated by  the  assumed dominance of abelian monopoles. 
     The classical field of an abelian monopole centered at $c$ is given by~\cite{Ripka:2003vv}   
     ( $x_n \equiv n\cdot x$, $x^i_\perp \equiv x^i - n^i x_n $), 
          \begin{equation}
     a^{ia}(x,\alpha) =  q a^i(x-c,n) w^a, 
     \end{equation} 
     where 
     \begin{equation} 
     a^i(x-c,n) = 
     \frac{g}{4\pi} \frac{ [(x -c) \times n]^i }{|x-c| [|x-c|  - (x_n-c_n)]}, 
  \end{equation}
   $\alpha = (q,c,n)$ denotes collectively the   monopole coordinates, 
     $q \pm 1$ is  the monopole charge in units of the  magnetic charge $g$, $c$ represents its location   and  $n$ is a unit vector that defines the orientation of the (straight) Dirac string. In the following we will neglect fluctuations in the relative orientation of the monopole  color orientations, 
   {\it i.e} in Eq.~(\ref{psit}) we set $w$ to be the common  orientation for all monopoles.  This restriction can be easily removed, however,  analytical calculations, even in the weak coupling limit that we discuss below, would not be possible. 
 The abelian component of the gauge field  $A_L$ is  then assumed to fluctuate over a background of 
      monopole-antimonopole  gas. For $N$ monopoles we thus write
      ($\alpha = (\alpha_1,\cdots \alpha_N)$),
  \begin{equation} 
    \Psi^{w,\alpha}_{L,N}[A_L] \propto   e^{-\frac{1}{2}\int dx dy A_L(x,\alpha) \omega_L(x-y)  A_L(y,\alpha)   },
    \end{equation}
where 
\begin{equation}
 A_L(x,\alpha) \equiv  A_L(x) - \sum_{i=1}^N a(x,\alpha_i).  \label{AL}
\end{equation} 
Finally, after summing over the monopole coordinates, averaging over their (common) 
  color orientation  and summing over the $N$ configurations we  obtain the ansatz for the vacuum wave functional given by,  
  \begin{equation} 
 |\Psi[A]|^2   =  \frac{1}{Z[0]} \int \frac{dw}{4\pi} D\alpha  \sum_{N=0}^{\infty}  
  \Psi_m(  \alpha)  | \Psi^{w,\alpha}_{L,N}[A_L]\Psi^w_T[A_T]|^2. \label{wfinal} 
\end{equation}  
The integration measure over monopole coordinates is given by $D\alpha \equiv 
   \Pi_{i=1}^N  \left[ dc_i  (dn_i/4\pi) (1/2   \sum_{q_i = \pm 1)}  \right] $. 
The distribution of monopoles is specified by the wave function $\Psi_m(\{\alpha\})$. We first discuss the non-interacting approximation, 
\begin{equation} 
 \Psi_m(\alpha)  =  \frac{\rho^N }{N!}  = const.,
 \end{equation} 
with $\rho$ being the density of monopole pairs.  The gluon propagator in this case is simply given by, 
\begin{eqnarray}
& & {\cal V}^{-1} \langle A^{ia}(k) A^{jb}(-k) \rangle =  \delta^{ij}_T(k) \delta^{ab} D(k), \nonumber \\
& & D(k) =  \frac{2}{3} D_T(k) + \frac{1}{3} D_L(k) 
\end{eqnarray}
with 
\begin{eqnarray} 
D_L(k) & = &  \frac{1}{2 \omega_L(k)} +  \rho \int \frac{dn}{4\pi} [ a^i(k,n) a^i(-k,n)]  \nonumber \\
&= & \frac{1}{2\omega_L(k)} + \frac{g^2 \rho}{2 k^4} \int_{-1}^1 dx \frac{1-x^2}{x^2 + \epsilon^2}.   \label{IRd} 
\end{eqnarray}
The  monopole  contribution is singular in the limit $\epsilon \to 0$ due to the collinear singularity associated with momentum component along the Dirac  string~\cite{Ripka:2003vv,Cornwall:2008da}. Even if this divergence was to be regularized, for example by combining monopole-antimonopole  paris into closed chains, the contribution remains strongly enhanced in the  IR  due to the $1/k^4$ behavior which originates from the long-range,  Coulomb, monopole field. Thus a simple model  with non-interacting monopoles (same result is obtained in the case of vortices) cannot be adequate since it gives as strongly enhanced gluon propagator that is  inconsistent  with all lattice results. 

The IR suppression of gluon propagator must therefore originate from screening by the interacting monopoles.  To this extent we introduce an effective interaction which is  repulsive (attractive)  between monopole and (anti) monopole, respectively, 
\begin{equation} 
V_{ij} = V(\alpha_i,\alpha_j) = q_i q_j \delta^2(n_i - n_j) V(c_i - c_j,n_i), \label{inter} 
\end{equation} 
and replace $\Psi_m$ by the corresponding partition function 
\begin{equation} 
\Psi_m(\alpha) = \frac{\rho^N}{N!} e^{-\frac{1}{4} \sum_{i,j=1}^N  V_{ij} }.
\end{equation} 
We choose the potential $V$ in Eq.~(\ref{inter}) in the form 
\begin{equation} 
V(k,n) = \int dx V(x,n) e^{ik \cdot x} =  \frac{4\pi M^{1+2 \gamma}}{k_n^2 k^{2+2\gamma}}. 
\end{equation} 
As will be seen below, the $1/k_n^2$, ($k_L = n \cdot k$) term will be responsible for screening in the direction  along the string white the critical exponent  $\gamma$ will  control the IR behavior of the propagator, and $M$ will related to the inverse Debye length.   

In calculation of  matrix elements, 
$ \langle {\cal O} \rangle \equiv \int  DA {\cal O}[A] |\Psi[A]|^2$ with the wave function given by 
Eq.~(\ref{wfinal}) 
 the summation over the number of monopoles  is done  with the help of  
   an   auxiliary   sine-Gordon field~\cite{m3},  $\phi(x,n)$.  In particular for a generating functional 
 \begin{equation} 
 Z[J] \equiv \int DA e^{\int dx [J^i_L(x) A^i_L(x) + J^{ia}_T(x) A^{ia}_T(x) ]  }| \Psi[A] |^2 , 
\end{equation} 
one finds 
\begin{equation} 
Z[J] = Z_T[J_T] \int D\phi e^{-S_L[\phi,J_L]} ,
\end{equation} 
with $Z_T[J_T] = \exp(  \frac{1}{4} \int dx dy J_T(x) \omega^{-1}_T(x-y) J_T(y) )$ and 
\begin{eqnarray} 
&&  S_L[\phi,J_L] =  \frac{1}{4\pi M^{1+2\gamma}}
 \int dx dn  \left[ \phi(x,n)  \partial^2_L (\partial^2)^{1+\gamma}  \phi(x,n) \right.  \nonumber \\
& & \left.  +  \lambda_D^{-(4 + 2\gamma)}  
[ 1 - \cos(  \phi(x,n) - v_L(x,n) ) ] \right] , \label{SL}
\end{eqnarray}
where 
\begin{equation} 
v_L(x,n) = i \int dy J^i_L(y) a^i(y-x,n). 
\end{equation} 
We defined  the Debye screening length $\lambda_D = (M^{1 + 2\gamma} \rho)^{-1/(4 +2 \gamma)} $.   In terms of the generating functional the gluon propagator is given  by $ \partial^2 \ln Z[J=0]/\partial_{J(k)}  \partial_{J(-k)}$.  
 In the mean-field approximation, which is valid in the limit $\lambda_D >  \rho^{-1/3}$, 
  where  $\rho^{-1/3}$ is the average separation between the monopoles,  the functional integral over $\phi$ can be computed analytically using the 
  saddle point approximation. At this level it is clear that interaction between monopole pairs screens their charges and thus regularizes the IR divergent part of the gluon propagator, leading to a propagator which is of the form  given by Eq.~(\ref{IRd}) but with the last term on the right hand side
   replaced by 
   
\begin{eqnarray}   
& &   \frac{g^2 \rho }{2 } 
    \int_{-1}^1 dx (1-x^2)\frac{ k^{2\gamma } }{
   k_n^2 k^{2+2\gamma}  + \lambda^{-4-2\gamma} }  = \nonumber \\ 
& & = k_D^{2\gamma} (g^2 \rho \lambda^4_D)  \left[   
    \frac{  (1 + k_D^4) \arctan(k^2_D) - k^2_D  }{k_D^6} \right], \label{screen} 
\end{eqnarray} 
where  $k_D \equiv k \lambda_D$.  In ~\cite{Burgio:2008jr} lattice simulation of the Coulomb propagator in $D=4$ was performed and   shown to be well approximated by the Gribov formula, $D(k) = k/(2 \sqrt{k^4 + M^4} )$  with $M= 0.890\mbox{ GeV}$. In ~\cite{Greensite:1979yn,Greensite:2009mi,Greensite:2010tm} it was  shown that the $D=3$ YM action in Landau gauge  may be a good approximation to $D=4$ Coulomb gauge wave  function, and recently   the corresponding Coulomb propagator was evaluated and shown to compare favorably with the exact  $D=4$ propagator~\cite{Quandt:2010yq}. For comparison we  then use results obtained with the $D=3$ YM action since it allows to separate the non-magnetic and magnetic contributions.  
  In  Fig.~1 circles represent the Coulomb propagator from ~\cite{Quandt:2010yq} with vortices removed. We use this propagator to fix $\omega_L$ and $\omega_T$ under a simplifying assumption $\omega_L(k)  = \omega_T(k)$ which we parametrize as  $(k/m)^a/2 ((k/m)^2 + b^2)^{1/2 + a/2}$ so that at low momentum $\omega_{L,T} \sim k^a$ and in the UV, $\omega_{L,T} \sim 1/k$.  The  fit of  $\omega$ to the propagator from ~\cite{Quandt:2010yq} with vortices removed yields,  $a= 1.21$, $b=0.29$, $m=3.88\mbox{ GeV}$. The result of the fit 
    is shown by the dashed line. 
   We then fix the three  parameters that describe our propagator ({\it c.f.}   Eqs.~(\ref{IRd}),~(\ref{screen})),  $L \equiv  g^2 \rho \lambda^4_D$,   $\lambda_D$ 
     and $\gamma$ by fitting the full gluon propagator from~\cite{Quandt:2010yq} (squares in Fig.~1) 
 and obtain $\lambda_D = 1.51 \mbox{ GeV}^{-1}$,  $ L = 6.46\mbox{ GeV}^{-1}$ and $\gamma=0.32$.  The result is shown by the solid line. And the  condition of applicability of the
    mean-field approximation requires week  monopole-monopole interaction ({\it } i.e. strong chromoelectric coupling, $e = 4\pi/g$)  with 
   \begin{equation} 
   g < \sqrt{\frac{L}{\lambda_D}} \lsim  2.
   \end{equation} 
\begin{figure}[h]
{\includegraphics[height=7cm]{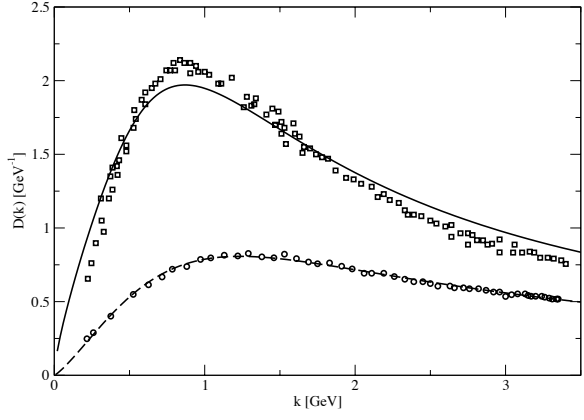}}
\caption{  Comparison of our  gluon propagator with that obtained from lattice computations
 ~\cite{Quandt:2010yq} } 
 \label{fig:1}
\end{figure}
   As discussed in Sec.~\ref{intro}, and shown above, while  the  IR suppression of the gluon might be the result of screening of magnetic charges,  IR enhancement of the ghost propagator comes from YM field distribution near the Gribov horizon. 
    We postpone a detailed numerical study of the ghost propagator    
     and  just notice   that because monopoles introduce orientation in color,  the diagonal  ($D_L$) and off-diagonal  ($D_T$) gluon propagators 
       are different,  and the mean-field  relation  between the ghost and gluon propagators  becomes more complicated.  In particular,  defining  the longitudinal and transverse ghost 
   form factors $d_L(k)$ and $d_T(k)$ by 
   \begin{eqnarray} 
 & &   \frac{d_L(k)}{k^2} \equiv {\cal V}^{-1}    \int DA w^a w^b   M^{-1}[A]    |\Psi^w[A]|^2 \nonumber \\
 & &  \frac{d_T(k)}{k^2} \equiv  {\cal V}^{-1} \int DA    \frac{\delta_T^{ab}(w)}{2}   M^{-1}[A]  
    |\Psi^w[A]|^2, \nonumber \\ \label{dtl} 
  \end{eqnarray}
 where  $M^{-1}[A] = M^{-1}[A](k,a,-k,b)=  -e (\nabla \cdot D[A])^{-1} $  is the inverse of the  Faddeev-Poppov operator. In the mean-field approximation one obtains,  
 \begin{eqnarray} 
& &  d^{-1}_L(k) = e^{-1} - I[D_T,d_T]  \nonumber \\
& & d^{-1}_T(k) = e^{-1} - \frac{1}{2} I[D_L,d_T] - \frac{1}{2} I[D_T,d_L],  \label{irnew} 
\end{eqnarray}
where ~\cite{swift} 
\begin{equation}
I[D,d] = \frac{N_C}{2} \int d\p D(p) \frac{ p k - p \cdot k}{p k  (p - k)^2} d(p-k).
\end{equation} 
 In the case $D_L = D_T = D$  these give $d_L = d_T = d$ with $d$ given by 
  \begin{equation} 
   d^{-1}_L(k) = e^{-1} - I[D,d]  \label{d} 
    \end{equation} 
 that has been studied in the past~\cite{Szczepaniak:2001rg,Szczepaniak:2003ve,Feuchter:2004mk,Epple:2006hv,Epple:2007ut}. 
  In general, after averaging over monopole color directions, 
  \begin{equation} 
  d(k) = \frac{2}{3} d_T(k) + \frac{1}{3} d_L(k). 
  \end{equation}   
  The IR  analyst  of Eq.~(\ref{d}) leads  to a relation between gluon and ghost critical exponents. That is assuming the IR behavior, of the form 
    $D(k) \propto k^{2\gamma}$ and $d(k) \propto k^{2\delta}$ one finds,  $\delta = -1/4 - \gamma/2$. Thus an  IR enhanced ghost propagator ($\delta < 0$) necessitates a  screened and IR suppressed ($\gamma > 0$) gluon propagator. 
  In  the case of Eq.~(\ref{irnew}), with $D_{L(T)}(k) \propto k^{2\gamma_{L(T)}}$ and $d_{L(T)}(k) \propto k^{2\delta_{L(T)}}$, respectively,  one finds $\delta_T = -1/4 -\gamma_L/2$ and $\delta_L = -1/4 - (\gamma_T - \gamma_L/2)$. In addition one should consider the Coulomb form factor and the gap equation. It was found  in ~\cite{Epple:2007ut} that  the coupled set of Dyson equations for these functions  admitted  only IR finite solutions. With the addition of  
    monopoles, however,  a  preliminary analysis of Eq.~(\ref{dtl})  indicates that IR critical solutions are possible.
    
    Finally we comment on the role of monopoles in suppressing large Wilson loop, 
    \begin{equation} 
    W_J[C] = \frac{1}{2J+1} Tr  \langle \Psi| P \exp( ie \oint_C dx^i A^{i,a} T^a ) |\Psi \rangle, \label{wl} 
    \end{equation} 
    where $T^a$ are the $SU(2)$ color generators in the $J$-th representation. 
    The integration over $A$  is computed by shifting $A_L$ according to   Eq.~(\ref{AL}). In the limit of large loops, the contribution from the non-monopole component (and $A_T$) is determined by  the gluon propagator, 
    \begin{eqnarray} 
\ln  W_J[C]  & \sim &  - \int_{C \to \infty}  dx^i  dy^j \delta^{ij}_T(\k)  \int \frac{dk}{(2\pi)^3} 
  D(k) e^{ik\cdot x}  \nonumber \\
   & \sim  & O( R^{-1} D(k \sim R^{-1}) , 
    \end{eqnarray} 
   where $R$ is the perimeter of the loop. An IR suppressed gluon propagator with $D(k) \sim k^{2\gamma}$ and $\gamma \ge 0$ leads to screening of the WIlson loop, {\it i.e.}  the loop  is dominated by shot range correlations and has at  most perimeter  dependence. 
   Thus if the long-range correlations are to dominate they must come from the monopole gas   
     and it is possible to ignore the  contributions from the fluctuating field. This leads to, 
             \begin{equation} 
   W_J[C]  =  \frac{1}{2J+1} \sum_{m = -J}^{J} \int D\phi e^{-S[\phi,m\eta[C]] } \label{www} 
   \end{equation} 
   where $S$ is given  by Eq.~(\ref{SL}) and $   \eta[C](c,n)  = \oint_C dy^i a^i(y - c,n) $. 
    Here $c$  is the location of a single monopole.  In particular for a large  loop  in $x-y$ plane with perimeter, $R  >> |x|$,  $\eta \to 2\pi \mbox{ sign}(c \cdot n)$. It immediately follows that $N$-ality zero ($J$-integer) loops are screened and all non-zero $N$-ality loops behave equivalently  to 
     the loop in the fundamental representation, $J=1/2$. The Casimir scaling presumably 
      comes from  the neglected  effects  of fluctuating field. 
  In the weakly-interacting limit   ($\lambda_D > \rho^{-1/3}$),
     the path integral in Eq.~(\ref{www}) can also be evaluated in the 
   saddle point approximation,  this time however the saddle point  does not correspond to 
   $\phi = 0$ but centers around $\eta$,~\cite{m3,stump} and is given by the solution of 
 \begin{equation} 
\frac{  (-\partial_L)^2 (-\partial)^{2 + \gamma}}{ \lambda_D^{-(4 + 2\gamma)} } \phi(x,n)  + \sin(\phi(x,n) + m \eta[C](x,n)) = 0. 
 \end{equation} 
 For large loops $R >> \lambda$ in the $x-y$ the equation becomes effectively one-dimensional , $\phi(x,n) \to \phi(z_n)$, $z_n = z \cdot n$ with $\phi(z)$ interpolating smoothly the 
 discontinuity in $\eta$ across the Wlson loop, 
  between $-\pi$ and $\pi$ over a distance of the order of 
   $\lambda_D$.  Substituting such a saddle point solution into the action in Eq.~(\ref{www}) 
    one finds 
   \begin{equation} 
   S[ \phi,\frac{1}{2} \eta ] \sim R^2 \lambda \frac{\lambda^{-4 + 2\gamma} }{M^{1 + 2\gamma}} 
 \end{equation} 
  and thus 
  \begin{equation} 
  \ln W_{1/2}[C] \sim  \lambda_D \rho A[C] 
  \end{equation} 
  where $A$ is the area of the loop. 
  
   \section{Summary and Outlook} 
\label{s2}

Since condensation of magnetic degrees of freedom is known to be present in the  QCD it becomes essential to include magnetic degrees of freedom 
  when constructing models of the QCD vacuum. In our construction we have assumed that these can be represented by a weakly-interacting gas of aligned (in color)  abelian  monopoles. Such a state is known to reproduce the confining properties for large, fundamental Wilson loops., however it does suffer from yielding $U(1)^N$ rather then $Z_N$ charge  dependence.   The latter could originate from  vortex configurations of long monopole chains and a construction of an ansatz for the corresponding vacuum sate would be highly desirable.  Here we have shown how magnetic monopoles influence the gluon propagator and have argued that the  IR suppression is the result of screening of magnetic charge and not confinement. Suppression of the  low momentum gluon propagator leads to suppression of long-range gluon fluctuations and restricts the field variables to the inside of the Gribov region. A self consistent calculation of gluon and ghost propagators and the Coulomb form factor is currently underway.

\section{ACKNOWLEDGMENTS  } 
We would like to thank Jeff Greensite and Hugo Reinhardt for numerous discussions. 
This work was supported in part by the US Department of Energy grant under 
contract DE-FG0287ER40365. We also acknowledge partial support  
 from the Australian Research Council Linkage International grant ÓLX0776452Ó that allowed his collaboration to continue.

\end{document}